# Deep-ultraviolet transparent conducting SrSnO$_3$ via heterostructure design

Fengdeng Liu[1,a*], Zhifei Yang[1,2,a*], David Abramovitch[3], Silu Guo[1], K. Andre Mkhoyan[1], Marco Bernardi[3,4], and Bharat Jalan[1,a]

[1]Department of Chemical Engineering and Materials Science, University of Minnesota – Twin Cities, Minneapolis, Minnesota 55455, USA

[2]School of Physics and Astronomy, University of Minnesota – Twin Cities, Minneapolis, Minnesota 55455, USA

[3]Department of Applied Physics and Materials Science, California Institute of Technology, Pasadena, California 91125, USA

[4]Department of Physics, California Institute of Technology, Pasadena, California 91125, USA

* Equal contributing authors
[a] Corresponding author: bjalan@umn.edu, liu00492@umn.edu, yang7001@umn.edu

**Abstract**

Exploration and advancements in ultra-wide bandgap (UWBG) semiconductors are pivotal for next-generation high-power electronics and deep-ultraviolet (DUV) optoelectronics. A critical challenge lies in finding a semiconductor that is highly transparent to DUV wavelengths yet conductive with high mobility at room temperature. Here, we achieved both high transparency and high conductivity by employing a thin heterostructure design. The heterostructure facilitated high conductivity by screening phonons using free carriers, while the atomically thin films ensured high transparency. We utilized a heterostructure comprising $SrSnO_3$/La:$SrSnO_3$/$GdScO_3$ (110) and applied electrostatic gating to effectively separate electrons from their dopant atoms. This led to a modulation of carrier density from $10^{18}$ cm$^{-3}$ to $10^{20}$ cm$^{-3}$, with room temperature mobilities ranging from 40 to 140 cm$^2$V$^{-1}$s$^{-1}$. The phonon-limited mobility, calculated from first principles, closely matched experimental results, suggesting that room-temperature mobility could be further increased with higher electron density. Additionally, the sample exhibited 85% optical transparency at a 300 nm wavelength. These findings highlight the potential of heterostructure design for transparent UWBG semiconductor applications, especially in deep-ultraviolet regime.

**Introduction**

The deep-ultraviolet (DUV) transparent conducting oxides (TCOs) are gaining attention as transparent electrodes for DUV light emitting diodes (LEDs)[1], with applications in biomolecule sensors[2], aerospace UV detectors[3], and pathogen control for food safety and water disinfection[4]. Among suitable materials, Al-doped ZnO (AZO) and Sn-doped $In_2O_3$ (ITO) have been used extensively as TCOs. However, their band gaps, $E_g \sim 3.3$ eV for AZO[5] and $E_g \sim 3.7$ eV for ITO[6] limit performance in the deep UV region (wavelength $\leq 300$ nm, corresponding to photon energy $\geq 4.1$ eV). In contrast, $SrSnO_3$ (SSO) emerges as a promising candidate for DUV applications[7] due to its wider bandgap of 4.1-4.6 eV[8-11] and high conductivity ($\sigma$) of ~3000 S cm$^{-1}$,[11] surpassing other well-known DUV TCOs such as α-$Ga_2O_3$ ($E_g \sim 5.3$ eV, $\sigma \sim 130$ S cm$^{-1}$)[12,13], β-$Ga_2O_3$ ($E_g \sim 5.0$ eV, $\sigma \sim 2500$ S cm$^{-1}$)[14], Al-doped $Mg_{0.43}Zn_{0.57}O$ ($E_g \sim 4.2$ eV, $\sigma \sim 407$ S cm$^{-1}$)[15] in terms of conductivity. Similar to widely studied $BaSnO_3$ (BSO), the conduction band of SSO derives from Sn 5s orbitals, which results in a low effective mass of electrons $m^* = 0.3 - 0.4$ $m_e$, where $m_e$ is the electron mass[16,17]. Such low electron effective mass combined with weak electron-phonon interactions has led to room-temperature mobilities as high as 56 cm$^2$ V$^{-1}$ s$^{-1}$ in *n*-doped SSO films grown by pulsed laser deposition (PLD), and as high as 70 cm$^2$ V$^{-1}$ s$^{-1}$ in films grown using hybrid molecular beam epitaxy (MBE)[18,19]. Truttmann et al. attribute the limited room-temperature mobility in hybrid MBE grown Nd-doped SSO films to phonon scattering, ionized-impurity scattering, and magnetic-impurity scattering owing to Nd[19]. Thus, to advance the room-temperature mobility of SSO and assess its suitability for transparent DUV devices, extrinsic factors such as impurities which limit the electron mobility must be addressed.

In this paper, we adopt a structure consisting of 4 nm SSO/19 nm La-doped SSO grown on $GdScO_3$ (GSO) (110) substrates by hybrid MBE. Assuming a doping density $N_B^{3D}$ for the La-doped

SSO (La:SSO) layer, we anticipate electron spillover when it forms a junction with the undoped SSO layer, as depicted in the top panel of Fig. 1(a). To illustrate this further, we show in Fig. 1(b) a normalized electron density profile $n^{3D}/N_B^{3D}$ across the sample structure by solving Poisson's equation with Thomas-Fermi approximation[20,21]. Here, $d = 0$ corresponds to the La:SSO/substrate interface, whereas $d = 19$ nm represents the SSO/La:SSO interface. It can be seen that an electron spillover from La:SSO to SSO is expected at the SSO/La:SSO interface owing to the Fermi level equilibration. In other words, this structure facilitates electron doping of undoped SSO near the interface thus separating carriers from their dopants. Additionally, the electron density in the undoped SSO layer can be adjusted either by modifying the chemical doping in the doped SSO layer (refer to the top panel in Fig. 1(a)), or via electrostatic gating[22,23] of undoped SSO (refer to the bottom panel in Fig. 1(a)). Using these two approaches, we successfully separate electrons from their donors, resulting in a remarkable modulation of electron densities ranging from $10^{18}$ cm$^{-3}$ to $10^{20}$ cm$^{-3}$ while achieving room-temperature electron mobilities as high as 140 cm$^2$V$^{-1}$s$^{-1}$, which is a two-fold increase over the previously achieved value in doped SSO. Quantitative analysis through first-principles calculations of electron–phonon interactions and phonon-limited mobilities at room temperature are consistent with our experimental findings while providing insights into the origin of high mobility. We further achieve a high optical transparency exceeding 85% at a 300 nm wavelength, establishing SSO as a promising UWBG semiconductor and a potential contender for emerging DUV transparent conducting oxide semiconductors.

We first discuss the structural characterization of 4 nm SrSnO$_3$/19 nm Sr$_{1-y}$La$_y$SnO$_3$/GdScO$_3$(110) samples as a function of doping concentration $y$. Details of the hybrid MBE growth process are provided in the Methods Section. The doping concentration $y$ was controlled by varying the La cell temperature $T_{La}$ from 1080 °C to 1250 °C at a fixed beam equivalent pressure (BEP) ratio of Sr/Sn. The high-resolution X-ray diffraction (HRXRD) coupled $2\theta$-$\omega$ scans of these samples in Fig. 2(a) show clear Laue oscillations, which demonstrates that the films are uniform with a high structural quality. These results give out-of-plane lattice parameters ranging from 4.10 Å to 4.12 Å, consistent with the strain-stabilized tetragonal SSO polymorph on GSO (110)[17]. The *in-situ* reflection high-energy electron diffraction (RHEED) patterns of these films after growth along the $[1\bar{1}0]_{orth}$ direction in Fig. 2(b) exhibit tall streaks and bright half-order streaks which indicate smooth surfaces and high crystalline quality, consistent with the HRXRD results. Previous study of La-doped SSO thin films grown by the same method has demonstrated a one-to-one correspondence between La dopant concentration and measured carrier density, indicating full activation of the La dopants[24]. To investigate La dopant activation in the films of this study, we utilize the Arrhenius equation to plot the measured 2D carrier concentration from Hall measurement $n_H^{2D}$ at room temperature on a log-scale as a function of inverse $T_{La}$ in Fig. 2(c). We find that for $T_{La} \leq 1240$ °C, $\log(n_H^{2D})$ vs. $1/T_{La}$ follows a linear behavior, which suggests a complete dopant activation[19,25]. For $T_{La} > 1240$ °C, the measured 2D carrier density decreases despite the film being grown at the highest $T_{La}$, suggesting that there is a solubility limit above $T_{La} = 1240$ °C. For films grown at $T_{La} < 1080$ °C, two-terminal resistance measurements show resistance values greater than 200 MΩ and no measurable conduction is observed. To confirm the crystallinity and interface structures of such thin films, high-angle annular dark-field scanning transmission electron microscopy (HAADF-STEM) images are obtained. A cross-sectional atomic-resolution HAADF-STEM image in Fig. 2(d) (see Methods) shows a representative 4 nm SSO/19 nm La: SSO/GSO (110) sample along the $[1\bar{1}0]_{orth}$ zone axis, which confirms an epitaxial and coherent interface between SSO and GSO substrate. No interface is observed between the top 4 nm undoped SSO and 19 nm La:SSO layers. A low magnification HAADF-STEM image is shown in Supplementary Fig. S1, which demonstrates no measurable defects across the whole SSO/La:SSO film except some intermixing over 2-3 unit cells at the interface between La:SSO and the GSO substrate.

As shown in Fig. 1, we anticipate an internal charge transfer from the doped SSO layer to the undoped SSO layer. An indirect indication of this transfer is the observed conduction in all samples during two-terminal measurements, where multimeter probes contact the surface. In the absence of charge transfer or electron doping in the undoped layer, the surface measurements of the films are expected to display insulating behaviors. Note that the undoped SSO surface without a buried doped SSO layer remains insulating.

To experimentally validate the potential charge transfer in SSO/La:SSO, we fabricate a series of superlattice structures consisting of $x$ nm SSO/(2.7-$x$) nm La:SSO ($x$ = 0, 0.6, 1.1, 2.1), totaling 7 layers, on GSO (110) substrates, as illustrated in Fig. 2(e)-(f). The La cell temperature (1240°C) and the total film thickness are kept constant for all samples. In Fig. 2(g), we plot the measured sheet conductance $\sigma_{2D}$ ($\sigma_{2D} = 1/R_s$, and $R_s$ is the measured sheet resistance) as a function of thickness $x$ at room temperature. Assuming the conductivity of La-doped SSO $\sigma_{3D}$ at room temperature is constant for all films due to the same doping density, the sheet conductance $\sigma_{2D}$ of these films, without considering any charge transfer, can be described by $\sigma_{2D} = \sigma_{3D} \times 7 \times (2.7-x)$, shown in Fig. 2(e)-(f) and represented by the solid straight line in Fig. 2(g), where only the doped layers contribute to the overall conductance.

However, the measured conductance $\sigma_{2D}$ deviates from a straight line as $x$ increases, suggesting that the conductance depends not only on the La:SSO layers. One possibility is that electrons diffuse from the doped layer into the undoped layer, leading to the observed change in conductance. These observations thus suggest an internal charge transfer consistent with our expectation illustrated in Fig. 1.

**Modulation of electron density in SrSnO₃ layer.**

We now turn to the discussion of the transport properties of 4 nm SrSnO$_3$/19 nm Sr$_{1-y}$La$_y$SnO$_3$/GdScO$_3$(110) samples as a function of doping concentration $y$. Fig. 3(a) shows the temperature-dependent sheet resistance, $R_s$. As the La cell temperature decreases, $R_s$ vs. $T$ reveals a metal-to-insulator transition, marked by the crossing of quantum resistance ($h/e^2$ ~25.8 kΩ where $h$ is the Planck's constant, and $e$ is the electron charge). An upturn in $R_s$ at low temperature for $T_{La}$ = 1120 and 1150°C samples is observed, which is attributed to the weak localization effect and electron-electron interactions[26]. Figure S2 shows results from the temperature-dependent Hall measurements.

In Fig. 3(b), we show the Hall mobility $\mu_H$ (defined as $\frac{1}{R_s n_H^{2D} e}$, where $R_s$ is the sheet resistance, $n_H^{2D}$ is the 2D Hall density, and $e$ is the electron charge) as a function of $n_H^{2D}$ at 300 K and 50 K, revealing a positive correlation, i.e. $\mu_H$ increases with increasing $n_H^{2D}$. Once again, this observation suggests the dominant electrical conduction is not in the doped SSO layer, but rather it is in the undoped layer due to charge transfer. If electrical conduction was confined to the doped SSO layer (i.e. no charge transfer), one would expect a decrease in mobility with increasing doping density due to ionized impurity scattering, and this impurity scattering should become even stronger at low temperatures owing to the suppression of phonon scattering processes. This is exactly opposite to the observation in Fig. 3(b). This result is consistent with our discussion in Figs. 1(b) and 2(g) that the transport is occurring in both channels. The transport properties of the undoped SSO layer can be separated from the La:SSO layer by applying a discrete two-channel conduction model[27]. In this model, the film with thickness $d$ can be separated into two layers, the accumulation layer and the bulk layer, where the conductivity and Hall coefficient are given by[27]

$$n_H^{2D}\mu_H = n_A^{3D}\mu_A d_A + n_B^{3D}\mu_B d_B \quad (1)$$

$$n_H^{2D}\mu_H^2 = n_A^{3D}\mu_A^2 d_A + n_B^{3D}\mu_B^2 d_B \quad (2)$$

where $n_H^{2D}$ and $\mu_H$ are the measured Hall sheet carrier density and Hall mobility, and $n_i^{3D}$, $\mu_i$ are effective 3D carrier density and mobility of the accumulation layer (i = A) and the bulk layer (i = B) with a thickness $d_i$. Note that $d_A + d_B = d$, the total thickness of the film. Equations (1) and (2) can thus be reduced to

$$n_H^{2D}\mu_H = n_A^{2D}\mu_A + n_B^{2D}\mu_B \quad (3)$$

$$n_H^{2D}\mu_H^2 = n_A^{2D}\mu_A^2 + n_B^{2D}\mu_B^2 \quad (4)$$

where $n_A^{2D}$ and $n_B^{2D}$ are sheet carrier density in the accumulation and bulk layers respectively.

We are interested in the mobility of the SSO layer which we will treat as the accumulation layer due to the accumulation of spilled electrons where $d_A = 4$ nm. The La-doped SSO layer will be assumed as the bulk layer where $d_B = 19$ nm. To extract the accumulation-layer mobility, one must first determine the electron densities in the bulk layer $n_B^{2D}$ and in the accumulation layer $n_A^{2D}$ respectively. One may think of using a 19 nm La-doped SSO without the 4 nm undoped SSO layer grown on GSO (110) to calibrate the doping density and mobility in the bulk layer, but it has been shown that electrons in doped SSO thin films may be compensated by surface depletion and strong localization effect due to surface disorders[18]. To avoid any bias, we use a self-consistent method to estimate the range of the bulk-layer properties and then extract accumulation-layer properties, as illustrated in Fig. 3(c). We assume the sheet electron density in the bulk layer $n_B^{2D}$ is within a fraction $\pm f$ of the measured $n_H^{2D}$ since 19 nm La:SSO likely contributes significantly to the measured 2D carrier density as $d_B \gg d_A$. Experimentally, we have also grown a 19 nm La-doped

SSO films on GSO (110) with the same conditions as the 4 nm SSO/19 nm La-doped SSO film where $T_{La}$ is kept at 1210 °C for both films. Measured room-temperature sheet carrier densities $n_H^{2D}$ are $2.28\times10^{14}$ cm$^{-2}$ for the 19 nm La:SSO film and $2.34\times10^{14}$ cm$^{-2}$ for the 4 nm SSO/19 nm La:SSO film, which are within ~2% difference. We then use Poisson's equation with Thomas-Fermi approximation to retrospectively calculate the doping density in the bulk layer $N_B^{3D}$, which in turn yields a 3D electron distribution depth profile $n^{3D}(z)$ across the structure where the sheet carrier density in the bulk layer $n_B^{2D} = \int_0^{d_B} n^{3D} dz$ and in the accumulation layer $n_A^{2D} = \int_{d_B}^{d_B+d_A} n^{3D} dz$ can be extracted. In Fig. 3(d), we show the extracted $n_B^{2D}$ and $n_A^{2D}$ as a function of the doping density $N_B^{3D}$ assuming the fraction factor $f = \pm 2\%$ based on the comparison above. Note that $n_B^{2D} < N_B^{3D} \times t_{doped}$, as expected due to some charges spilling over and $n_A^{2D}$ is one order of magnitude less than $n_B^{2D}$. With these values, we extract the accumulation-layer mobility $\mu_A$ as a function of 3D electron density $n_A^{3D} = n_A^{2D}/d_A$ at 300 K and 200 K by solving the equations (3) & (4), which are shown in Fig. 3(e) and (f). The green shaded areas indicate the overall range of $\mu_A$ extracted based on $f = \pm 2\%$. $\mu_A$ increases with increasing $n_A^{3D}$ and with $n_A^{3D}$ ~$5.6\times10^{19}$ cm$^{-3}$, $\mu_A$ reaches around 139 cm$^2$V$^{-1}$s$^{-1}$ at 300 K for the $T_{La} = 1240$°C sample at lower bound and over 200 cm$^2$V$^{-1}$s$^{-1}$ at the higher bound, at least 55% improvement from the measured mobility $\mu_H$. Similar improvement is also shown at 200 K. The high mobility in the accumulation layer at 300 K is likely due to the absence of charged impurities defects in the 4 nm SSO layer. If this were to be true, a similarly high mobility should also be obtained using electrostatic gating[28]. To this end, we perform an electrolyte gating experiment where the carriers in the 4 nm accumulation layer are varied by applying an external bias.

**Electrostatic gating of SSO/La:SSO thin film by ion gels.**

Electrolyte gating with ionic liquids and ion gels in electric double-layer (EDL) transistor configurations can induce high electric fields at the electrolyte/solid interface and carrier modulation up to $10^{15}$ cm$^{-2}$ in oxide thin films can be achieved[22,23]. A Hall bar device is first fabricated on the 4 nm SSO/ 19 nm La-doped SSO film grown on GSO (110) substrate where the La cell temperature is 1210 °C. 7 nm Ti capped with 30 nm Au is deposited as metal contacts and coplanar gate electrodes. An optical image of the device is shown in Fig. 4(a). Then an ion gel is placed on the sample covering the channel area and gate electrodes. A schematic of the device with the ion gel illustrating the formation of electrical double-layer at the ion gel/SSO interface is shown in Fig. 4(b). A reversible electrostatic window with applied gate voltage $V_g = \pm 4V$ is determined by monitoring the sheet resistance when charging and discharging the EDL capacitor, as detailed in Supplementary Figs. S3-S4. The sample is first measured with positive gate voltages up to +4 V. Then a second measurement with another piece of ion gel with negative gate voltages down to -3 V is performed after the sample is cleaned with acetone and isopropyl alcohol in an ultrasonication cleaner. Measured 2D carrier density $n_{\text{Hall}}^{2D}$ and Hall mobility $\mu_{\text{Hall}}$ as a function of temperature at gate voltages from -3 V to +4 V are shown in Fig. 4(c) and (d) respectively. Measurement without ion gel done on a 19 nm La-doped SSO thin film grown on GSO (110) substrate under the same growth conditions with $T_{\text{La}} = 1210$ °C is also shown for comparison. Note here that the difference of $n_{\text{Hall}}^{2D}$ in both films at 250 K falls within 2%. In Fig. 4(c), $n_{\text{Hall}}^{2D}$ increases with positive $V_g$ due to electron accumulation, and $n_{\text{Hall}}^{2D}$ decreases with negative $V_g$ as expected due to electron depletion. In Fig. 4(d), $\mu_{\text{Hall}}$ also increases with positive $V_g$, but $\mu_{\text{Hall}}$ remains relatively constant at negative $V_g$. The electrostatic modulation due to electrolyte gating is typically confined within a few nanometers from the electrolyte/film interface[29,30]. This modulated layer can in principle possess different transport properties than the bulk part of the film, and the measured values do not reflect this layer accurately[29,30]. The properties of the modulated layer can be separated from the bulk layer by applying the same discrete two-channel conduction model listed in equations (3) & (4). The bulk region with thickness $d_B$ is usually assumed to maintain the identical properties of those at $V_g = 0$ V; however, such assumption is usually applied when modelling electrostatic modulation on a uniformly doped thin film[29,30], and at $V_g = 0$ V in our structure, there are already two conduction layers as discussed in the last section due to charge transfer. Therefore, to accurately extract the accumulation-layer properties due to electrostatic gating, we do not apply the same assumption. Here, the region near the ion gel/SSO interface which is modulated by the ion gel is the accumulation layer and the rest of the sample is the bulk region. With the application of a negative $V_g$, negative ions will accumulate in the ion gel at the ion gel/SSO interface, and electrons in the SSO layer due to charge spillover from La:SSO will be depleted. This is evident by the similar value of $n_{Hall}^{2D}$ at $V_g = -1$ V with measured sheet carrier density of the comparison sample, 19 nm La-doped SSO grown on GSO (110), shown in Fig. 4(c). Effectively, the measured mobility at negative $V_g$ should reflect the intrinsic mobility of the bulk region of the sample without contributions from the electron accumulation in the undoped SSO layer. The modest $V_g$-induced change of $\mu_{Hall}$ at positive $V_g$ and relatively no change at negative $V_g$ can be explained due to the large thickness of the film or the high chemical doping in La:SSO ($> 1\times10^{20}$ cm$^{-3}$). We can then consider that the measured $n_{Hall}^{2D}$ and $\mu_{Hall}$ at $V_g = -1$ V represents the sheet electron density $n_B^{2D}$ and mobility $\mu_B$ in the bulk region of the film with thickness $d_B$. In the presence of electrostatic gating with ion gels, the bulk region is assumed to retain these transport properties. Therefore, to extract the accumulation-layer electron density $n_A^{2D}$ and mobility $\mu_A$ due to the electrostatic modulation, we need to estimate the thickness of the accumulation region $d_A$.

The accumulation-layer thickness $d_A$ can be determined through charge density profile. The depth-dependent electron density distribution $n^{3D}(d)$ at various positive $V_g$ at 250 K can be obtained by solving Poisson's equation with Thomas-Fermi approximation by treating electrons in the accumulation layer as 3D electron gas and with appropriate boundary conditions[21,29,30], as shown in Fig. 4(e). The $d = 0$ corresponds to La:SSO/substrate interface, while $d = 19$ nm corresponds to the SSO/La:SSO interface and $d = 23$ nm is the ion gel/SSO interface. In the bulk of the film, away from the SSO/La-doped SSO interface, $n^{3D}$ approaches the chemical doping level calculated using the same procedure outlined in Fig. 3(c). At the SSO/La-doped SSO interface, $n^{3D}$ changes due to the charge transfer from doped SSO to undoped SSO by Fermi level equilibration. Toward the ion gel/La:SSO interface, $n^{3D}$ increases with increasing $V_g$ due to the induced electrons by gating. The value of accumulation-layer thicknesses $d_A$ can be obtained from the $n^{3D}$ profiles and we define it as the thickness where 90% of electrostatically induced electrons are confined within[29,30]. (See the Supplementary Information Section V for the calculation and the error analysis on the accumulation-layer parameters). The extracted $d_A$ as a function of $V_g$ at 250 K is plotted in the inset of Fig. 4(e). $d_A$ decreases with increasing $V_g$, due to the increasing confinement by gate-induced electric fields at the ion gel/SSO interface. $d_A$ values are then used to calculate $n_A^{3D}$ ($n_A^{2D}/d_A$) and $\mu_A$ at different gate voltages, shown in Fig. 4(f). Note that $d_A$ values are between 3.3 nm and 3.5 nm, which is just slightly less than the thickness of the undoped SSO layer, 4 nm. This self-consistently confirms that the extracted values through this modeling will reflect the electron doping of the undoped SSO layer. $n_A^{3D}$ increases with increasing $V_g$ because higher gate voltages induce more electrons in the accumulation layer. The total induced 2D electron density $\Delta n^{2D}$, which is the integrated difference between the electron density at an applied gate voltage and the electron density at 0 V over the accumulation layer thickness $d_A$, starts with

~$8\times10^{12}$ cm$^{-2}$ at $V_g = +1$ V and increases monotonically with $V_g$, reaching ~$2.3\times10^{13}$ cm$^{-2}$ $V_g = +4$ V. This is in the same order of magnitude as reported values in ionic-liquid and ion gel gated semiconductors[22,30]. The value of $\mu_A$ reaches around 140 cm$^2$V$^{-1}$s$^{-1}$ at $V_g = +1$ V at 250 K and decreases as $V_g$ increases. Scattering due to the ion gel at the ion gel/SSO interface can limit the mobility in the accumulation layer, and the accumulation-layer thickness decreases with increasing $V_g$ due to high confinement which will limit the mobility as $V_g$ increases[28]. Nevertheless, $\mu_A$ is ~70% to ~100% improved compared to $\mu_H$ ($V_g = 0$ V), and more than 168% increase compared to the sample without the 4 nm SSO layer on top. This value from the electrostatic gating is again consistent with the values observed in Figure 3(e) determined from the 4 nm SSO/ 19 nm La:SSO heterostructure, reinforcing the reliability and robustness of these findings. Finally, to further confirm these results, we perform transport calculations using the first-principles.

**First-principles phonon-limited mobility calculations.**

Since there are no dopants in the undoped SSO layer, the ionized impurity scattering should not play a role in limiting the mobility. Therefore, the question becomes what limits the electron mobility in undoped SSO layer. We carry out first-principles calculations of the phonon-limited mobilities[31] with the 3D carrier concentrations ranging from $5\times10^{18}$ cm$^{-3}$ to $1\times10^{20}$ cm$^{-3}$ at temperatures larger than 200 K, since phonon occupations increase at higher temperatures. A detailed analysis of electron-phonon interactions in the orthorhombic phase of SSO has been discussed in our previous work[19]. In this study, all films exhibit strain-stabilized tetragonal structure, so we perform calculations in the tetragonal structure. Fig. 5(a) shows the phonon dispersion curves in tetragonal SSO along several high-symmetry lines in the Brillouin zone. The color scale on the phonon dispersion curves indicates the electron-phonon coupling strength between different phonon modes and the electronic state at the conduction band minimum $|g_v(k = \Gamma, q)|$. Fig 5(b) shows the band structure of the conduction band used in the calculations.

Using these interactions, we compute the phonon-limited mobility, including scattering from all phonon modes on the same footing. In the calculations, the carrier density can be modified by tuning the chemical potential. Note here that the calculated mobility is drift mobility. To confirm it is reasonable to compare the calculated mobilities with the experimental values which are Hall mobilities, we also calculate the Hall factor $\mu_{Hc}/\mu_d$, where $\mu_{Hc}$ is the calculated Hall mobility and $\mu_d$ is the drift mobility, at selected carrier concentrations $n^{3D} = 1\times10^{19}$ cm$^{-3}$, $5\times10^{19}$ cm$^{-3}$, $1\times10^{20}$ cm$^{-3}$ between 200 K and 350 K, shown in Fig. 5(c). In the range of carrier concentrations and temperatures shown, the Hall factor is nearly 1, so it is reasonable to consider drift mobility as an approximation of Hall mobility.

Fig. 5(d) shows the plot computed phonon-limited mobility at 200 K and 300 K for 3D carrier concentrations $n^{3D}$ ranging from $5\times10^{18}$ cm$^{-3}$ to $1\times10^{20}$ cm$^{-3}$. The calculated phonon-limited mobility increases with carrier concentration and reaches values of 118 cm$^2$V$^{-1}$s$^{-1}$ at 300 K and 263 cm$^2$V$^{-1}$s$^{-1}$ at 200 K for $n^{3D} = 5\times10^{19}$ cm$^{-3}$. These values are in very good agreement with our experimental mobility range of 115– 166 cm$^2$V$^{-1}$s$^{-1}$ at 300 K and 260– 289 cm$^2$V$^{-1}$s$^{-1}$ at 200 K for an electron density of ~5.4 ×10$^{19}$ cm$^{-3}$ (see Fig. 3(e) and Fig. 3(f)). We do not show calculations below 200 K as the electron-phonon scattering is suppressed at lower temperatures and is not the dominant mobility-limiting mechanism. We also compare the extracted accumulation-layer mobility from the electrolyte gating at 250 K with the calculated phonon-limited mobility, as shown in Supplementary Fig. S5. The extracted accumulation-layer mobility is shown to be smaller than the calculated phonon-limited mobility. Note that scattering from the ion gel at the ion gel/SSO interface may limit the mobility. This effect is not included in our calculations and may lead to small discrepancies between the extracted and computed mobilities. However, the relatively good match between our experimental data and the independent theoretical calculations gives us confidence that the mobility values we extracted truly reflects the mobility of the electron-doped SSO layer, which is mainly limited by electron-phonon scattering above 200 K and at room temperature due to the absence of charged defects.

**New outlook and opportunities.**

There are two classes of materials being proposed as transparent conductor candidates, correlated metals and doped wide-bandgap semiconductors. In Fig. 6(a), we compare the achieved electrical conductivity for some of the well-known transparent conducting semiconductors as a function of their bandgaps[7,11,13-15,32-42] along with the conductivity of the correlated metals SrNbO$_3$[43,44] and SrVO$_3$[45]. In our sample, the effective conductivity across the whole structure (4 nm SSO/19 nm La:SSO) exceeds 4900 S cm$^{-1}$, which we include in Fig. 6(a) for comparison. Among materials with ultra-wide bandgaps ($E_g$ > 4.1 eV) including MgZnO[15], Ge$_x$Sn$_{1-x}$O$_2$[32], Al$_x$Ga$_{1-x}$N[38,40], α-Ga$_2$O$_3$[13], β-Ga$_2$O$_3$[14], and ZnGa$_2$O$_4$[42], our SSO film exhibits the highest conductivity with a total thickness of 23 nm. SrNbO$_3$ has been demonstrated to exhibit electrical conductivities over 26,000 S cm$^{-1}$ in epitaxial thin films[43,44]. In addition to the electrical conductivity, transparency in the deep-ultraviolet (DUV) region is also important to evaluate the performance of DUV transparent conductors. Fig. 6(b) shows the optical transmittance at a wavelength near 300 nm (the boundary of the DUV region) for the same or similar films used in Fig. 6(a) except the α-Ga$_2$O$_3$[46], AlN[47] and Al$_x$Ga$_{1-x}$N[48,49] films. Note here that the optical transmittance of SrNbO$_3$ is calculated from the absorption coefficient obtained through spectroscopic ellipsometry[43] instead of a value obtained from direct measurements. Among the materials plotted in Fig. 6(b), SrSnO$_3$[7,11], α-Ga$_2$O$_3$[46], AlN[47], Al$_x$Ga$_{1-x}$N ($x \geq 0.7$)[48] and SrNbO$_3$[43] exhibit high transmittance over 70%. For comparison, we also include the optical transmittance of our 3 nm SSO/13 nm La-doped SSO film grown on a double-side-polished GSO (110) substate with $T_{La}$ = 1145 °C, which shows an 85% transmittance at λ = 300 nm. The measured optical transmittance as a function of wavelength of this sample is plotted in the Supplementary Fig. S6(a). The relatively high optical transmittance compared to previous reports[7,11] can be attributed to the lower thickness of the sample structure. Nevertheless, it shows that SSO heterostructure exhibits both the highest DUV transparency and highest conductivity among all-known doped wide-bandgap semiconductors.

As illustrated in Figs. 6(a) and 6(b), due to the relatively low conductivity of α-Ga$_2$O$_3$, AlN and Al$_x$Ga$_{1-x}$N ($x \geq 0.7$), only SrSnO$_3$ and SrNbO$_3$ stand out with both high conductivity and high transmittance at λ ~ 300 nm. The high conductivity in the correlated metal SrNbO$_3$ thin films is achieved with a carrier concentration of ~10$^{22}$ cm$^{-3}$ along with an average mobility of ~8 cm$^2$V$^{-1}$s$^{-1}$ at room temperature[43]. To compare the mobility, we plot the extracted room-temperature mobility in electron doped SSO as a function of the electron density along with other TCOs in the Supplementary Fig. S7. Electron-doped SSO exhibits a room-temperature mobility of 139 cm$^2$V$^{-1}$s$^{-1}$ at an electron density ~5.6×10$^{19}$ cm$^{-3}$. This electron mobility is nearly twice the highest value (~70 cm$^2$V$^{-1}$s$^{-1}$)[18,19] reported to date in SSO and ~10 times that of SrNbO$_3$[43]. Electron-doped SSO also maintain higher mobilities with similar carrier concentrations with α-Ga$_2$O$_3$[13] and β-Ga$_2$O$_3$[50]. These results place electron doped SSO on par with many other DUV TCOs in terms of mobilities.

In summary, we grow a series of the 4 nm SSO/19 nm La:SSO thin films with a range of La doping levels. We identify the charge transfer from La:SSO into SSO, and separate electrons from the dopant atoms. By changing the doping density in the La-doped SSO layer and by electrostatic gating with ion gels, we are able to control the electron density in the SSO layer. The combination of controlled doping, detailed transport measurements with modeling and first-principles calculations of electron-phonon interactions allowed us to understand, tune and predict the transport behaviors and mobility-limiting mechanisms in electron-doped SSO. The experimentally extracted mobility in the electron-doped SSO thin film from 200 – 300 K is close to the value predicted by the theory for an ideally pure SSO crystal. These results combined with high optical transparency at 300 nm wavelength and high electrical conductivity establish SSO as a highly promising deep-ultraviolet transparent conductive oxide in optoelectronics.

**Methods.**

**Growth and characterization of SSO thin films.** All the SrSnO$_3$ thin films were grown by a radical-based hybrid molecular beam epitaxy method[51,52]. All films were grown on GdScO$_3$ (110) single-crystal substrates at a fixed substrate temperature of 950°C. The substrates were cleaned *in-situ* with oxygen plasma for 25 minutes prior to growth, ensuring no carbon contamination at the surface. The oxygen flow was set to 0.7 standard cubic centimeters per minute (sccm) to achieve an oxygen background pressure of 5×10$^{-6}$ Torr while applying 250 W of RF power to the plasma coil. The beam equivalent pressure (BEP) of Sr was fixed at 6×10$^{-8}$ mbar, as measured by a retractable beam flux monitor. The BEP ratio of chemical precursor hexamethylditin (HMDT, by which Tin was supplied) to Sr was kept at 550 to ensure stoichiometric growth. These conditions achieved a growth rate of ~33 nm per hour. Dopant density in the films was controlled by varying La cell temperature (1080 °C ≤ $T_{La}$ ≤ 1250 °C). The SSO/La:SSO superlattices with 7 iterations were grown on GSO(110) substrates with $T_{La}$ = 1240 °C and the thickness of undoped SSO was controlled by changing the growth time. A Rigaku SmartLab XE diffractometer was used for X-ray diffraction measurements. Film thicknesses were extracted from the X-ray reflectivity (XRR).

**Scanning transmission electron microscopy (STEM).** To prepare the cross-sectional lamellas for STEM study on a representative SSO/La-doped SSO/GSO (110) sample, Focused Ion Beam (FIB) (FEI Helios NanoLab G4 dual-beam) was used. An amorphous carbon layer of 75 nm thickness was deposited on the film using thermal evaporation. In addition, a protection layer of 1.5 μm amorphous C was deposited on the targeted area prior to ion beam trenching. FIB was operated at 30 kV. STEM experiments were conducted using an aberration-corrected FEI Titan G2 60–300 (S)TEM microscope equipped with a CEOS DCOR probe corrector, monochromator, and a super-X energy dispersive X-ray (EDX) spectrometer. STEM was operated at 200 kV. HAADF-STEM images were acquired with a probe current of 90 pA, camera length of 130 mm, and probe convergence angle of 25.5 mrad. The inner and outer angles of the detector were set to 55 mrad and 200 mrad, respectively.

**Transport measurement.** Electrical transport measurements were performed in a Dynacool physical property measurement system (PPMS) (Quantum Design) in a Van der Pauw geometry with aluminum wire bonding to connect samples to the resistivity measurement puck. The temperature range used in the measurement is between 1.8 K and 300 K and the magnetic field was swept between ±9T.

**Electrolyte gating experiments.** A Hall bar device with channel length 1 mm and width 0.5 mm was first patterned on the 4 nm SSO/19 nm La-doped SSO/GSO (110) sample and the 19 nm La-doped SSO/GSO (110) sample with $T_{La}$ = 1210 °C and similar growth conditions, using a hard mask and reactive-ion etching. Etching was performed in a reactive-ion etcher (Advanced Vacuum) with a mix of 40 sccm $SF_6$ and 10 sccm Ar with 300 W power in 9 mTorr pressure. 7 nm Ti capped with 30 nm Au was deposited on the devices using another hard mask for metal contacts and gate electrodes. Ion gels were prepared by mixing poly vinylidene fluoride-*co*- hexafluoropropylene (PVDF-HFP) polymer, 1-ethyl-3-methylimidazaolium bis-trifluoromethylsulfonyl imide (EMI/TFSI) ionic liquid, and acetone as the solvent in a ratio of 1:4:7. The solution was then spin-coated on a glass wafer at 500 rpm for 1 minute. The obtained gels were then heated at 70 °C for 24 hours inside a vacuum oven to remove residual solvent. Then an ion gel was cut by a blade and placed on the channel area and on the gate electrodes. Immediately after, the sample was transferred into the PPMS under a He environment to avoid potential degradation of the gels in the air. Gate voltage was applied, and leakage current was monitored using a Keithley 2634B sourcemeter. See Supplementary Information Section III & IV for discussion of the electrostatic window and leakage current at 250 K along with the reversibility of ion-gel gating. A DC voltage was first applied at 280 K and after waiting for >20 minutes to give ions sufficient time to move and charge the EDL capacitor, the sample was cooled down to 1.8 K. Transport measurements were performed while warming up in the Dynacool PPMS. The sample was first measured with positive gate voltages. Then a second run with another piece of ion gel with negative gate voltages down to -3V was performed after the sample was taken out of the PPMS and cleaned with acetone and isopropyl alcohol in an ultrasonication cleaner.

**First-principles calculation.** We performed ab initio calculations on SSO in the strained tetragonal I4/mcm phase with lattice parameters ($a = b = 5.615$ Å, $c = 8.310$ Å). The in-plane lattice parameters were set from experiment to match the GdScO$_3$ substrate[17], while the out-of-plane lattice parameter was relaxed to minimize energy. We computed ground state electronic structure with Density Functional Theory (DFT) and lattice dynamics and electron-phonon coupling with Density Functional Perturbation Theory (DFPT) using the Quantum ESPRESSO code[53]. All calculations used the PBEsol-GGA exchange-correlation functional and norm conserving pseudopotentials from PseudoDojo[54] using a plane-wave basis with a 90 Ry kinetic energy cutoff. The DFPT calculations computed electron-phonon coupling matrix elements on 8 × 8 × 6 electronic *k*-point and 4 × 4 × 3 phonon *q*-point coarse grids. We used the Perturbo code[31] for transport calculations using interpolations to fine *k*- and *q*-point grids of 120 × 120 × 90 (160 × 160 × 120 for Hall mobility) at temperatures above 100 K and 200 × 200 × 150 at 100 K and below. Phonon-limited drift and Hall mobility was computed by iteratively solving the Boltzmann transport equation with electron-phonon collisions[55].

We find an imaginary phonon mode in the phonon spectrum of the strained tetragonal SSO structure, which we neglect in transport calculations. This mode corresponds to a dynamic instability in the zero-temperature system and may be related to the lack of an explicit $GdScO_3$ substrate in the calculation or stabilization of the tetragonal phase at finite temperature. Such modes are common in higher symmetry phases of materials with complex structures and lattice dynamics, including many perovskite oxides, and are often neglected in transport calculations due to weak coupling to electrons compared to polar longitudinal optical modes[56,57].

**Optical property measurement.** The optical transmittance ($T$) of 3 nm SSO/13 nm La:SSO film grown on a double side polished GSO(110) substrate was measured by a spectrophotometer (Cary 7000, Agilent) over 210–500 nm wavelengths. The sample is mounted on a holder with a 5 mm diameter hole. The contribution of the GSO (110) substrate is identified by measuring its optical transmittance $T_0$ separately before growth. The optical transmittance of 3 nm SSO/13 nm La:SSO film is obtained by the ratio of two measurements $T/T_0$. Absorbance is calculated as the minus logarithm of the transmittance, and the absorption coefficient $α$ is calculated by accounting for the sample thickness.

**Acknowledgments**

Synthesis and characterization (F.L) were supported primarily by the Air Force Office of Scientific Research (AFOSR) through grants FA9550-21-1-0025 and FA9550-23-1-0247, and in part from the National Science Foundation (NSF) through award number DMR-2306273. Electrolyte gating (Z.Y.) was supported primarily by the UMN MRSEC program under Award No. DMR-2011401. Film growth was performed using instrumentation funded by AFOSR DURIP awards FA9550-18-1-0294 and FA9550-23-1-0085. S.G. and K.A.M. were supported partially by the UMN MRSEC and NSF award and No. DMR-2309431. Parts of this work were carried out at the Characterization Facility, University of Minnesota, which receives partial support from the NSF through UMNMRSEC. Portions of this work were carried out at the Minnesota Nano Center, which receives support from the NSF through the National Nanotechnology Coordinated Infrastructure (NNCI) under Award No. ECCS-2025124. DA acknowledges work supported by the National Science Foundation Graduate Research Fellowship under Grant No. 2139433.

**Author Contributions:** F.L., Z.Y., and B.J. conceived the idea and designed the experiments. F.L. grew the films. Z.Y. developed the fabrication process and performed gating experiments. F.L. and Z.Y., performed structural characterization and electrical testing. STEM/EDX measurements were performed by S.G. under the supervision of K.A.M. The first-principles calculations were performed by D.A. under the supervision of M.B. F.L., Z.Y., and B.J. wrote the manuscript. All authors contributed to the discussion and manuscript preparation. B.J. coordinated all aspects of the project.

**Competing Interest Statement:** The authors declare no competing interests.

**Data and materials availability:** All data needed to evaluate the conclusions of the paper are present in the paper and/or the Supplementary Materials.

**Supporting Information**

The supporting information contains additional details on the structural characterization using STEM of as-grown 4 nm SSO/19 nm La-doped SSO/GSO (110). Details of electrical transport measurements with and without gating plus the details of the calculations of accumulation-layer carrier density and mobility due to electrostatic modulation are included.

**Figures and Tables**

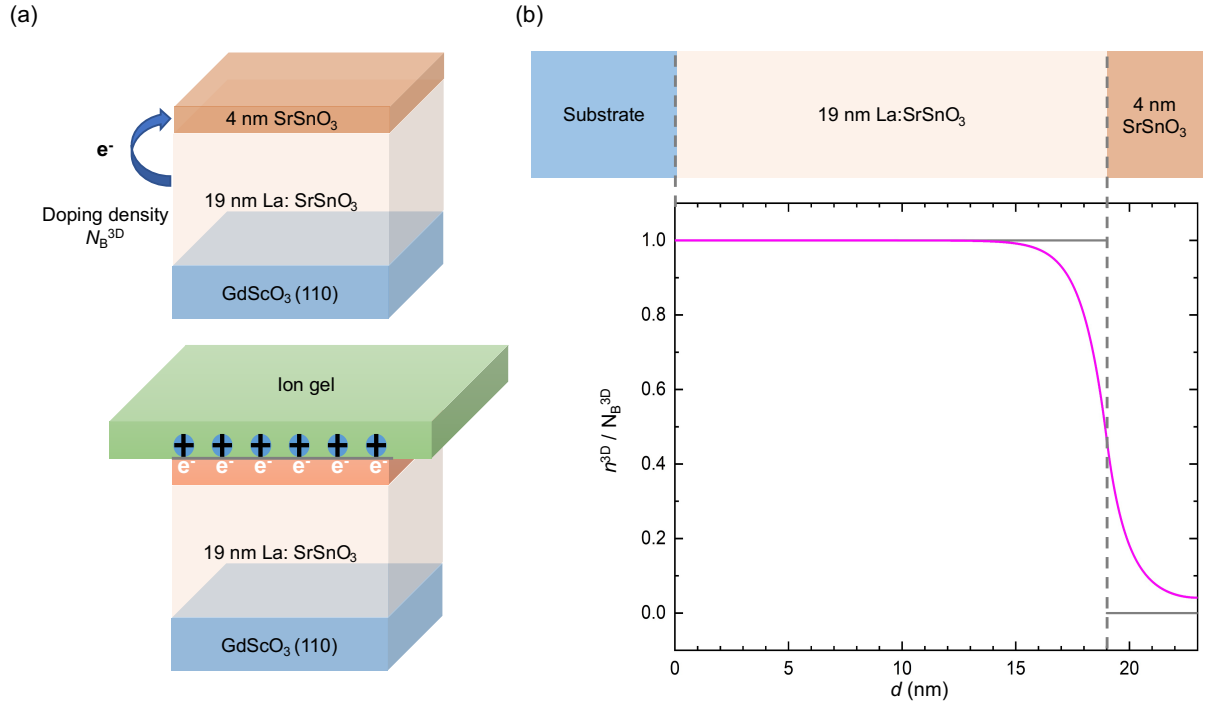

**Fig. 1. Methods of tuning the electron density in SrSnO₃ layers. (a)** Schematic of the sample structure. The top panel shows charge transfer at the SSO/La:SSO interface and modulation of electron density in the SSO layer can be achieved by changing the doping density in the La:SSO layer $N_B^{3D}$. The bottom panel shows the modulation of electron density in the SSO layer by electrolyte gating with ion gels. **(b)** Normalized 3D electron density $n^{3D}/N_B^{3D}$ across the sample by solving the Poisson's equation with Thomas-Fermi approximation with a schematic of the sample structure on top.

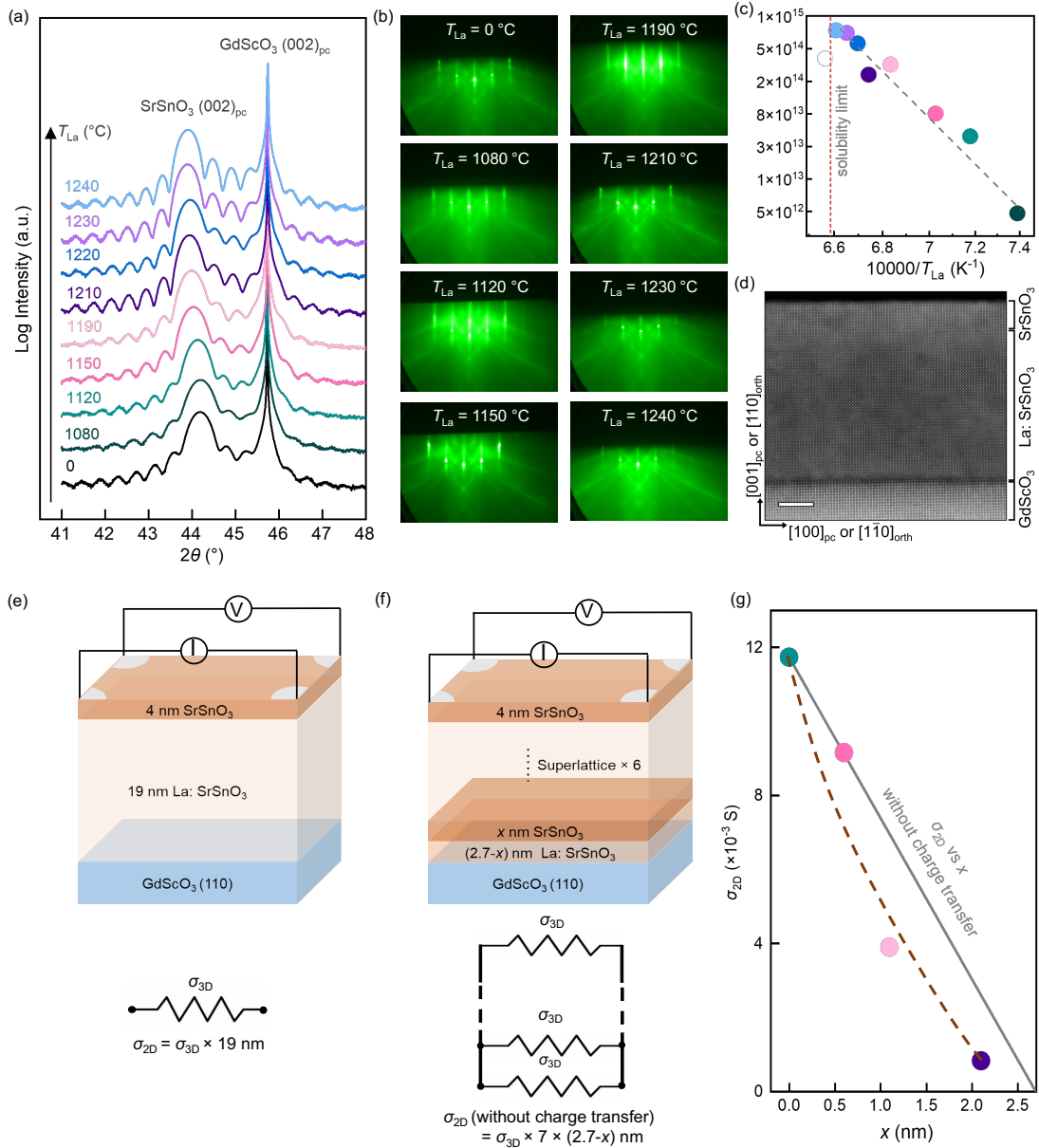

**Fig. 2. Characterization of SrSnO₃ thin films and the charge transfer. (a)** High resolution X-ray diffraction $2\theta$-$\omega$ couple scan of 4 nm SrSnO$_3$/19 nm La: SrSnO$_3$/GdScO$_3$ (110) thin films with different La cell temperatures. **(b)** Reflection high-energy electron diffraction (RHEED) patterns of these thin films with the beam pointing along the $[1\bar{1}0]_{orth}$ direction. **(c)** Measured 2D carrier concentrations $n_H^{2D}$ as a function of inverse of La cell temperatures $10000/T_{La}$. **(d)** Atomic resolution HAADF-STEM cross-sectional image of a representative 4 nm SrSnO$_3$/19 nm La: SrSnO$_3$/GdScO$_3$ (110) substrate. Scale bar = 5 nm. The image is low-pass filtered. **(e)** Schematic of the 4 nm SrSnO$_3$/19 nm La: SrSnO$_3$/GdScO$_3$ (110) structure and related conductance. **(f)** Schematic of the $x$ nm SrSnO$_3$/ (2.7 - $x$) nm La: SrSnO$_3$ superlattices structure, where $x$ = 0, 0.6, 1.1, 2.1, and relationship between measured conductance and conductivity. **(g)** Measured conductance $\sigma_{2D}$ as a function of SrSnO$_3$ layer thickness $x$. The solid line is the expected value of $\sigma_{2D}$ without charge transfer assuming a constant conductivity $\sigma_{3D}$. The dashed line is guidance for the eye.

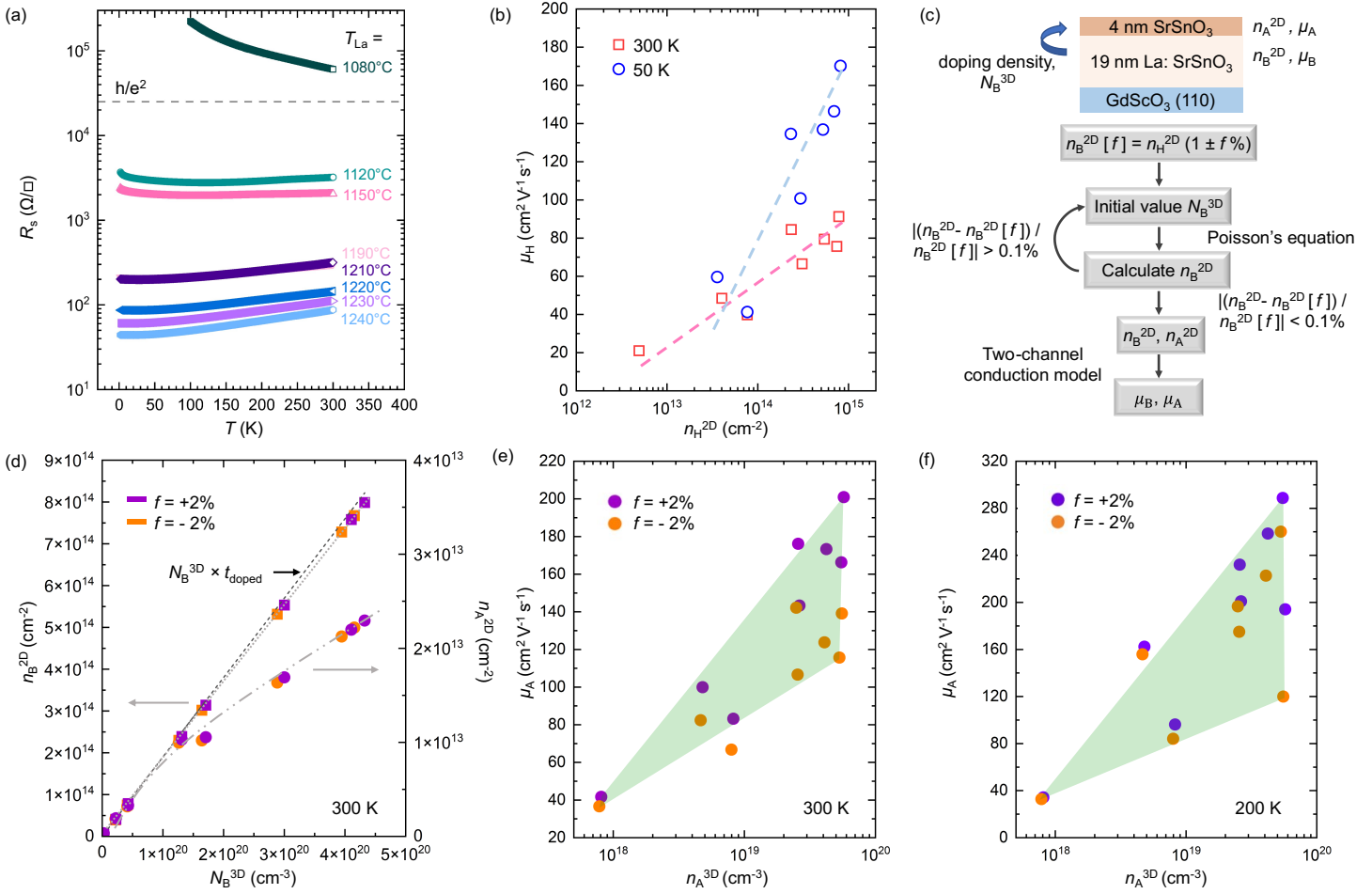

**Fig. 3. Modulation of electron density in the SrSnO₃ layer by tuning the doping density. (a)** Measured sheet resistance $R_s$ as a function of temperature for samples grown with $T_{La}$ ranging from 1080 °C to 1240 °C. **(b)** Measured Hall mobility $\mu_H$ as a function of sheet Hall carrier density $n_H^{2D}$ for the samples at 300 K and 50 K. **(c)** Self-consistent procedure to estimate the properties of the bulk and the accumulation layers. **(d)** Extracted sheet bulk carrier density $n_B^{2D}$ and accumulation carrier density $n_A^{2D}$ as a function of 3D doping density $N_B^{3D}$ with the fraction value $f = \pm 2\%$. Extracted accumulation-layer mobility $\mu_A$ as a function of accumulation-layer electron density $n_A^{3D}$ at **(e)** 300 K and **(f)** 200 K. The green shaded areas are for guidance to showcase the range of $\mu_A$.

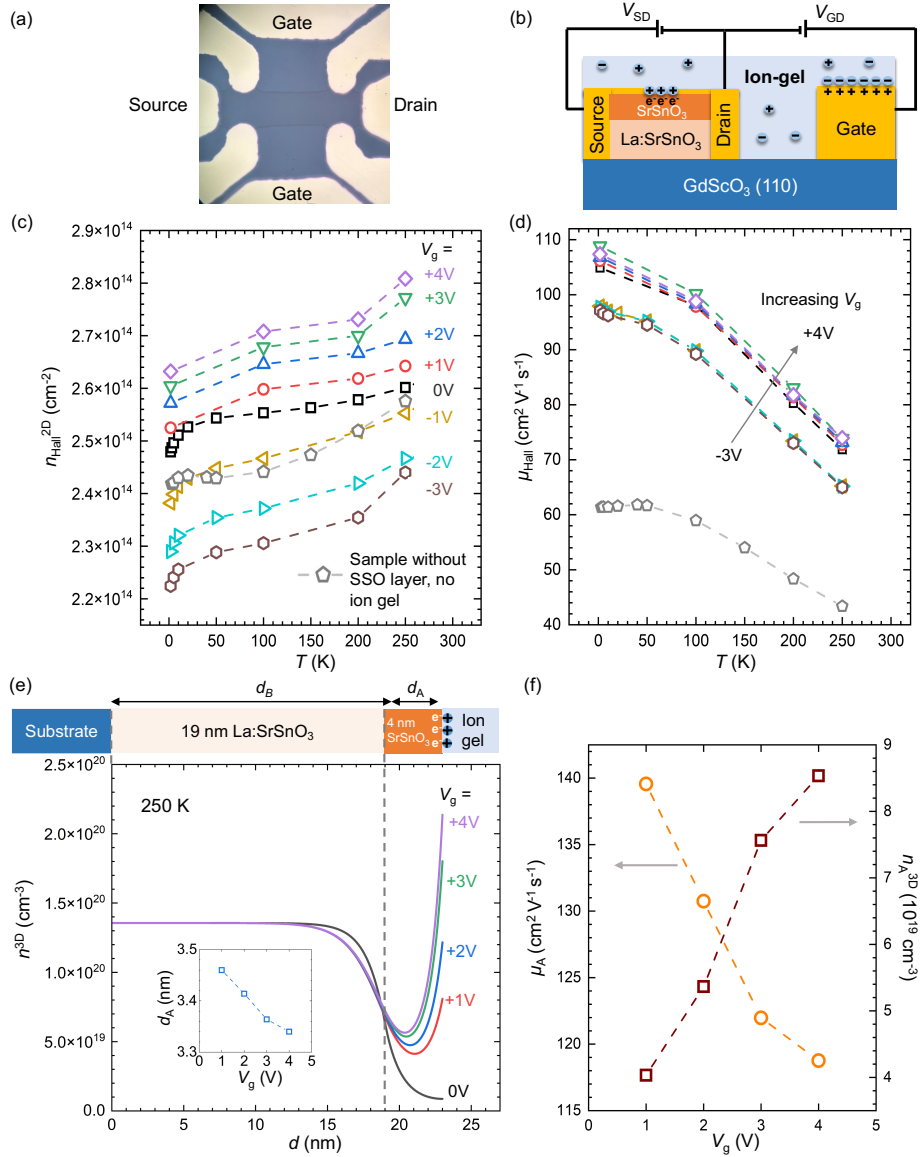

**Fig. 4. Electrostatic gating experiment with ion gels. (a)** Optical image of the device with source, drain and gate electrodes labeled. **(b)** Schematic of the device from the side view which shows the formation of electrical double layer at the ion gel/SrSnO$_3$ interface. **(c)** Temperature-dependent measured sheet carrier density $n_{\text{Hall}}^{2D}$ at gate voltages $V_g$ from -3 V to +4 V. The measurement on 19 nm La-doped SrSnO$_3$ grown with similar conditions is shown in the gray symbols. **(d)** Temperature-dependent measured mobility $\mu_{\text{Hall}}$ at gate voltages $V_g$ from -3 V to +4 V. **(e)** 3D electron density depth profile $n^{3D}$ vs. $d$ for various $V_g$ at 250 K with the schematic of the sample structure on top showing two-channel conduction. Inset shows the corresponding accumulation-layer thickness $d_A$ as a function of $V_g$. **(f)** The extracted $V_g$-dependence of accumulation-layer mobility $\mu_A$ and electron density $n_A^{3D}$.

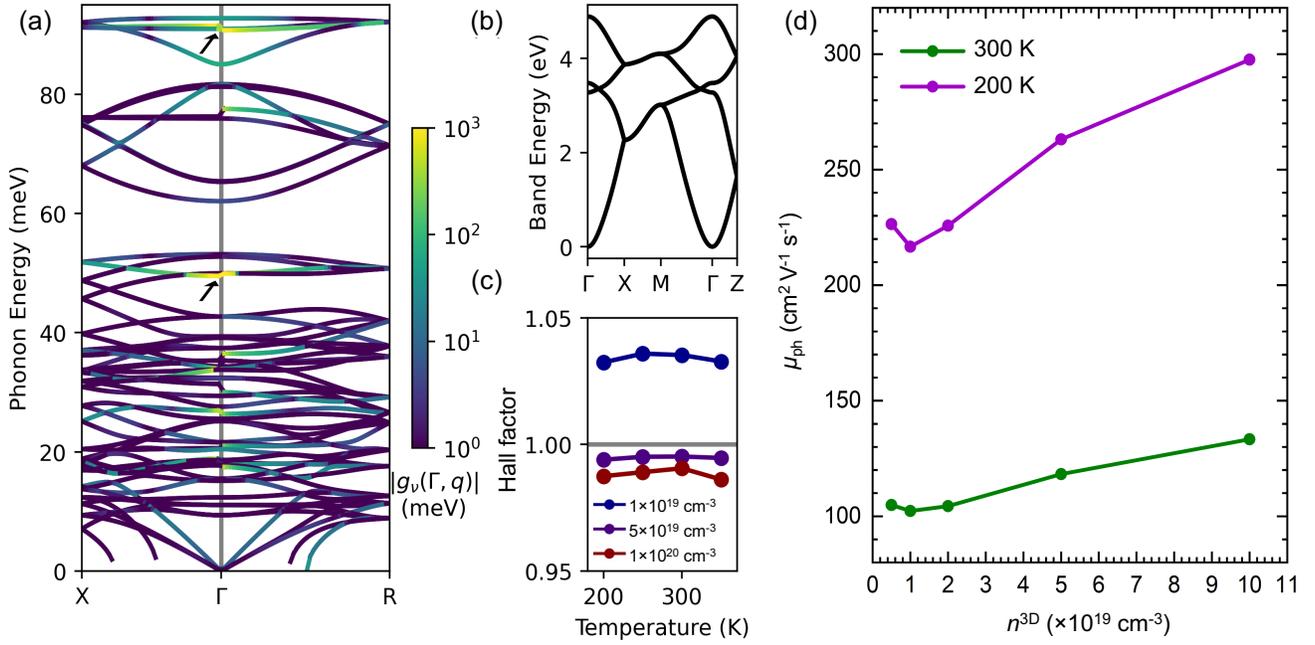

**Fig. 5. First principles calculations of electron-phonon limited carrier mobility in SrSnO₃.** (a) Phonon dispersions along high-symmetry lines in the Brillouin zone, color-coded according to the electron-phonon coupling strength between phonon each mode and the conduction band minimum $|g_v(k = \Gamma, q)|$. The arrows indicate strongly coupled longitudinal optical phonon modes. (b) Calculated band structure of the conduction band of SSO. (c) Calculated Hall factor $\mu_{Hc}/\mu_d$ for carrier concentrations $n^{3D}$ ranging from $1\times10^{19}$ cm$^{-3}$ to $1\times10^{20}$ cm$^{-3}$ as a function of temperature. (d) Phonon-limited electron drift mobility $\mu_{ph}$ as a function of carrier concentration $n^{3D}$ at 200 K and 300 K.

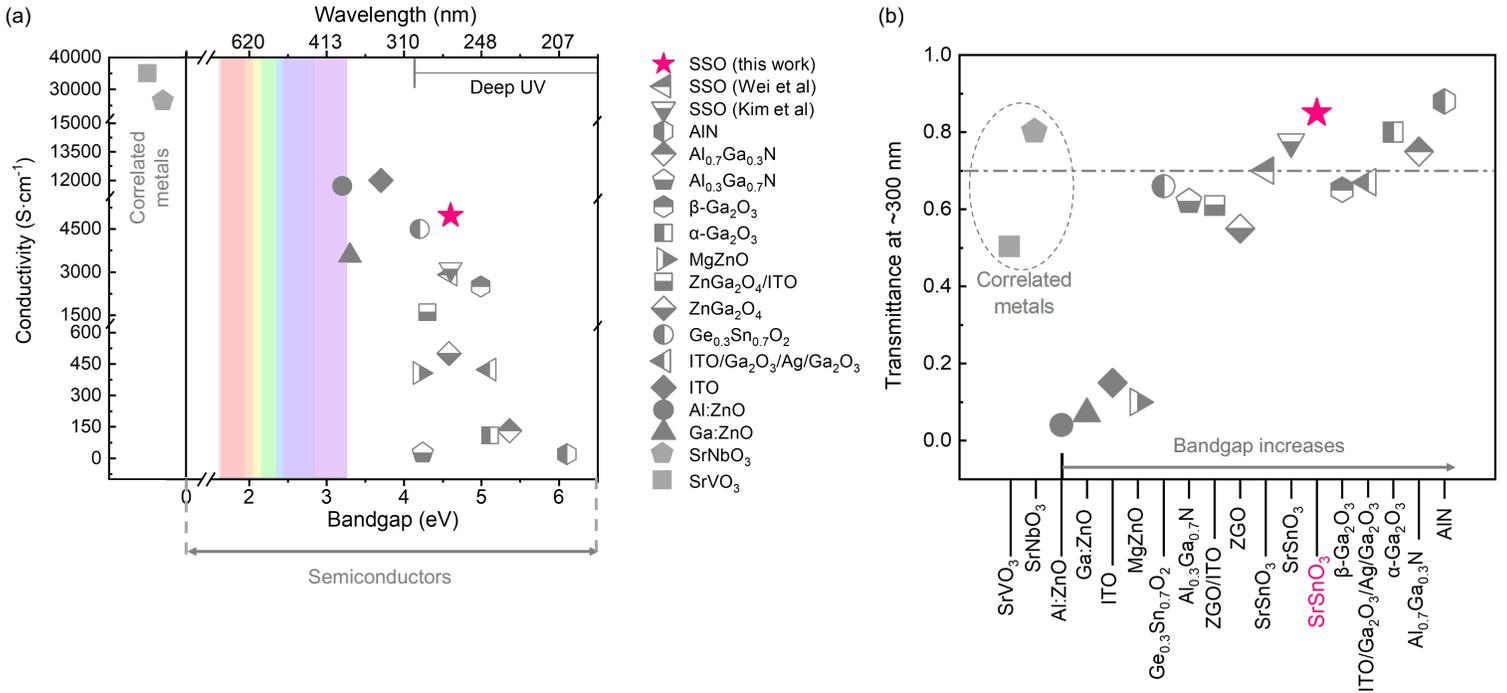

**Fig. 6. Comparison of the conductivity and optical transmittance of transparent conductors. (a)** Room-temperature electrical conductivity of selected transparent conducting materials as a function of the bandgap along with the conductivity of correlated metals SrNbO$_3$ and SrVO$_3$ thin films. The effective conductivity of 4 nm SrSnO$_3$/19 nm La: SrSnO$_3$/GdScO$_3$ (110) is included for comparison. **(b)** Optical transmittance at a wavelength ~300 nm for films outlined in (a). The thickness of the films from left to right: SrVO$_3$ (4 nm), Al: ZnO (280 nm), Ga: ZnO (180 nm), ITO (1000 nm), SrNbO$_3$ (10 nm), MgZnO (500 nm), Ge$_{0.3}$Sn$_{0.7}$O$_2$ (200 nm), Al$_{0.3}$Ga$_{0.7}$N (390 nm), ZnGa$_2$O$_4$/ITO (200 nm/25 nm), ZnGa$_2$O$_4$ (single crystal), SrSnO$_3$ (112 nm), SrSnO$_3$ (120 nm), SrSnO$_3$ (3 nm undoped/13 nm doped), β-Ga$_2$O$_3$ (200 nm), ITO/Ga$_2$O$_3$/Ag/Ga$_2$O$_3$ (10 nm/15 nm/7 nm/15 nm), α-Ga$_2$O$_3$ (600 nm), Al$_{0.7}$Ga$_{0.3}$N (518 nm), AlN (300 nm).